0341766/src> head Schwarz-Christoffel.tex
\documentclass[onecolumn, secnumarabic, amssymb, amsmath,nobibnotes, aps, prl,nofootinbib]{revtex4}

\usepackage{graphicx}

\date{october 21, 2011}

\newcommand{\half}{\mbox{\small{$\frac{1}{2}$}}}

\newcommand{\third}{\mbox{\small{$\frac{1}{3}$}}}

\begin{document}
\title{the lineage of string theory}
\author{B. H. Lavenda}
\email{info@bernardhlavenda.com}
\homepage{www.bernardhlavenda.com}
\affiliation{Universit$\grave{a}$ degli Studi, Camerino 62032 (MC) Italy}
\begin{abstract}
The Regge trajectories, upon which string theory is based, behave as rigid rotators rather than vibrating strings. The same relation, between the angular momentum, and the square of the mass, can be found in gravity, the electroweak, and strong interactions. The angle deficit for cosmic strings is shown to be an angle excess that is related to the increase of the circumference of a uniformly rotating disc.    Schr\"odinger's time independent equation with a centrifugal barrier gives an automorphic function that can be constructed as the ratio of its two independent solutions for values of the angular momentum lying outside of their positive, integer values.  If the fixed points $0$ and $\infty$ in the $z$-plane correspond to $-i$ and $+i$ in the $\omega$-plane, then elliptic substitutions tessellate the $\omega$-plane in the form of cresents, while if the fixed points correspond to $-1$ and $+1$ in the $\omega$-plane are source and sink, just like the lines of force between positive and negative charges. The unit disc undergoes a stretching by a Lorentz transform.
\end{abstract}
 
\maketitle
\section{Flexible Strings or Rigid Rotators?}
It is well-known that divergences in physical theories occur because of the approximation of treating particles as idealized points. Once their finite extension is accounted for, these divergences should be tamed. String theory compliments itself as being \lq\lq the first quantum theory, ever, to include gravity without also including ultraviolet divergences.\rq\rq~\cite{BZ} But, is this really the case?

Heisenberg realized that there were two types of \lq fundamental\rq\ constants: $\hbar$ and $c$, and $e$ and $G$. The latter pair contain the intrinsic coupling strengths to the electrodynamic and gravitational fields, the charge, $e$, and the Newtonian gravitational constant, $G$. The first pair of constants determine the  nature of the interaction; when Planck's constant, $\hbar$, appears, quantum effects are important, and when the speed of light appears $c$, relativistic effects occur.~\cite{BL} Now, a fundamental length, $\ell$, should be related to a fundamental mass, $m$, through the Compton relation, $\ell=\hbar/cm$, which is the linear dimension for constructing the best-localized state from a wave packet containing only plane-wave components. That such a distance cannot be determined from $e$, $\hbar$ and $c$ is intended to mean that electromagnetic interactions are carried by particles with no mass. We may say there is no effective range of interaction for electromagnetic waves, or that the range is infinite. Not so with gravity for a length can be determined from the combination $\surd(\hbar G/c^3)$, and, consequently, a mass, $\surd[(\hbar c)/G]$. This is reason enough to believe that gravity waves, if they exist, should not behave as electromagnetic waves and travel at speed $c$, even in the linear approximation to the gravitational equations of  general relativity. Moreover, the purported carriers of the gravitational interaction, gravitons, should likewise be limited in extent in contrast to weightless photons.

The reason why there are no divergences in string theory is because it introduces a single parameter, $\varepsilon$, which is the linear energy density. It is related to the string length $\ell_S$ by 
\begin{equation}
\ell_S=\hbar c\surd\alpha^{\prime}=\surd\left(\frac{\hbar c}{2\pi\varepsilon}\right), \label{eq:ellS}
\end{equation}
where
\begin{equation}
\alpha^{\prime}=\frac{1}{2\pi\varepsilon\hbar c}. \label{eq:alpha}
\end{equation}
The constant \eqref{eq:alpha} has been singled out because it is the slope of a Regge trajectory.
We will  show that $\varepsilon$ is analogous to $G^{-1}$, and to a whole slew of other coupling strengths. However, the association of an energy density with a string tension will be seen to be inappropriate.

\begin{figure}[t]
	\centering
		\includegraphics[width=0.3\textwidth]{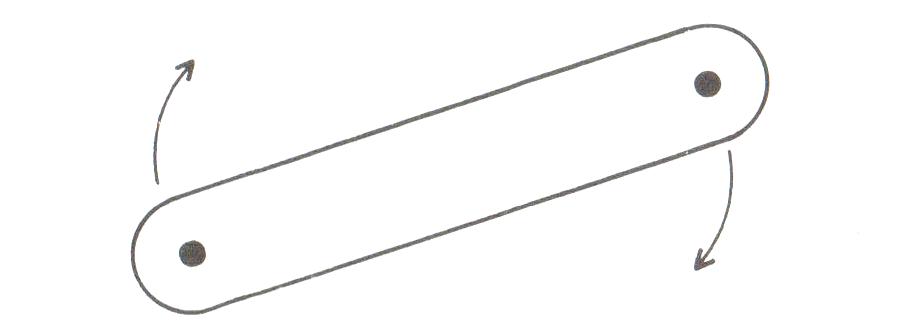}
	\caption{A rigidly rotating bar used to calculate the relation between the angular momentum and the mass of a hadron. The ends of the bar travel at the speed of light.}
	\label{fig:bar}
\end{figure}

\begin{figure}[b]
	\centering
		\includegraphics[width=0.5\textwidth]{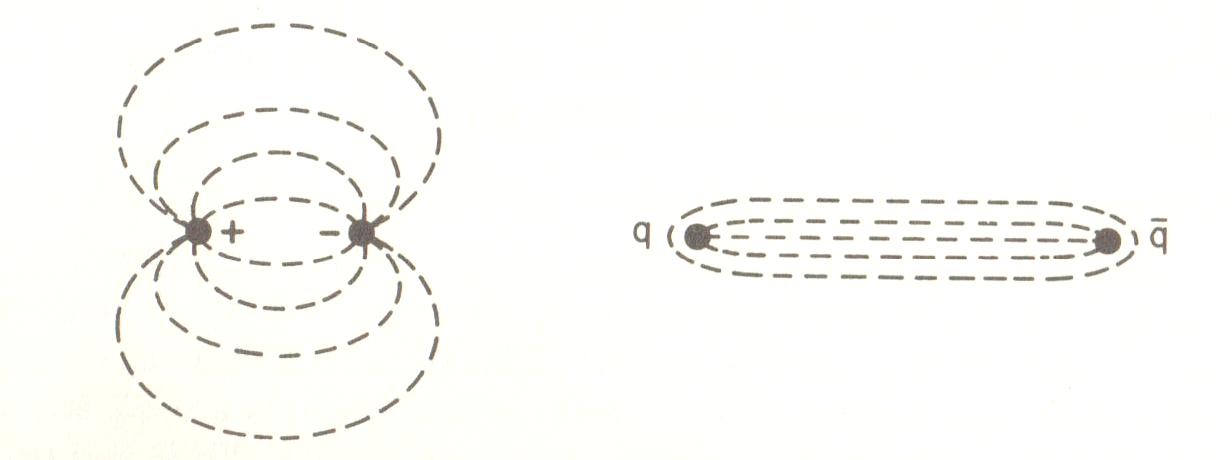}
	\caption{The analogy between the lines of force (left) between two opposite charges and the color lines of force between quark pairs (right). The lines are pictorially seen to  hold together the quarks by a tube or string on account of the strong self-interaction between gluons, the carriers of the color field.}
	\label{fig:string}
\end{figure}

Consider a \emph{rigid\/} rod of length $2r_0$, as shown in Fig.~\ref{fig:bar}. It is rather weird that this will become the model of a \emph{flexible\/} string, which is envisioned as connecting quark pairs, $q\bar{q}$, shown in Fig.~\ref{fig:string}. The rigid bar is free to rotate about its center of mass, and the velocity of its ends is $c$. Since we are considering a \emph{rigid\/} rotation, $r_0=c/\omega$, which defines the angular velocity $\omega$; for any other velocity, $r\omega<c$.

  The total string energy will be given by~\cite{DP}
\begin{equation}
E_S=2\int_0^{r_{0}}\frac{\varepsilon}{\surd(1-v^2/c^2)}dr=2\int_0^{c/\omega}\frac{\varepsilon}{\surd(1-r^2\omega^2/c^2)}dr. \label{eq:E}
\end{equation}
Now introducing a change of variable $r=(c/\omega)\sin\vartheta$ in \eqref{eq:E} results in
\begin{equation}
E_S=\frac{2c\varepsilon}{\omega}\int_0^{\pi/2}d\vartheta=\pi\varepsilon\frac{c}{\omega}. \label{eq:E-bis}
\end{equation}
The total angular momentum, $J$, in units of $\hbar$, can also be calculated relativistically, according to its definition
\begin{equation}
J_S=\frac{2}{\hbar c^2}\int_0^{r_{0}}\frac{\varepsilon r v}{\surd(1-v^2/c^2)}dr=\frac{2\varepsilon}{\hbar c^2}\int_0^{c/\omega}\frac{r^2\omega dr}{\surd(1-r^2\omega^2/c^2)} \label{eq:J}
\end{equation}
Again, introducing the change of variable into \eqref{eq:J} results in
\begin{equation}
J_S=\frac{2\varepsilon c}{\hbar\omega^2}\int_0^{\pi/2}\sin^2\vartheta\;d\vartheta=\frac{\pi}{2}\frac{\varepsilon c^2}{\hbar\omega^2}. \label{eq:J-bis}
\end{equation}

Now, \eqref{eq:E-bis} and \eqref{eq:J-bis} must be valid for whatever angular velocity we may choose. Eliminating $\omega$ from the two equations gives
\begin{equation}
J_S=\alpha^{\prime}E_S^2, \label{eq:Regge}
\end{equation}
where the constant of proportionality is given by \eqref{eq:alpha}. The Regge trajectory, \eqref{eq:Regge} has zero intercept. A non-vanishing intercept cannot be derived from classical phenomenologically. Since $v\propto r$, the rotation is rigid, and hardly resembles a flexible string. But, it does resemble other theories where there is but a single parameter that fixes the scale.

From \eqref{eq:E-bis}, the string mass may be simply written down as
\begin{equation}
M_S=\hbar\Big/c\surd(\hbar c/\varepsilon)=\surd\left(\frac{\varepsilon\hbar}{c^3}\right), \label{eq:MS}
\end{equation}
where we drop factors of $2\pi$ since we are interested only in the ratios of fundamental constants. In terms of \eqref{eq:MS}, the Regge trajectory, \eqref{eq:Regge} can be written as
\begin{equation}
J_S=\frac{M^2}{M_S^2}=\frac{G_S(Mc^2)^2}{\hbar c}, \label{eq:Regge-bis}
\end{equation}
where the \lq gravitational\rq\ coupling $G_S=1/\varepsilon$.  

The characteristic frequency of the string,
\begin{equation}
\omega_S=c/\ell_S=\surd\left(\frac{\varepsilon c}{\hbar}\right), \label{eq:omegaS}
\end{equation}
 is not the characteristic frequency of a real string~\cite[p. 75]{BZ},
\begin{equation}
\omega_S r=\surd\left(\frac{\varepsilon}{\varrho}\right)\label{eq:omegaS-bis}
\end{equation}
where $\varrho$ is the linear mass density, and $\varepsilon$ is the string tension [cf. \eqref{eq:string-bis} below].

It is already apparent that the so-called Planck scale inverts mass and length so that the mass and length are
\begin{gather}
M_G=\surd\left(\frac{\hbar c}{G}\right), \label{eq:MG}\\
\ell_G=\surd\left(\frac{G\hbar}{c^3}\right). \label{eq:ellG}
\end{gather}
It is apparent from \eqref{eq:MG} that we can write the Regge trajectory as
\begin{equation}
J_G=\frac{M^2}{M_G^2}=\frac{GM^2}{\hbar c}, \label{eq:ReggeG}
\end{equation}
which is a dimensionless ratio, just like the fine structure constant,
\begin{equation}
J_{\gamma}=\frac{e^2}{\hbar c}. \label{eq:FS}
\end{equation}
If \eqref{eq:FS} is truly angular momentum, then light has a rotational inertia of
\[
\mathcal{I}=\frac{1}{137}\frac{\hbar}{\omega}, \]
so that its angular momentum is
\[L=J_\gamma \hbar=\mathcal{I}\omega=e^2/c.\]

If \eqref{eq:ReggeG} is to hold on all scales it must hold on the Planck scale, which will serve as a check of self-consistency. Writing \eqref{eq:ReggeG} in the form
\begin{equation}
\ell_G^2\omega=\frac{GM_G}{c} \label{eq:ReggeG-bis}
\end{equation}
allows us to solve for the characteristic angular velocity. We thus find
\begin{equation}
\omega_G=\surd\left(\frac{c^5}{G\hbar}\right). \label{eq:omegaG}
\end{equation}
This could have been obtained directly from \eqref{eq:ellG}, i.e., $\omega_G=c/\ell_G$. Finally, the rotational inertia is
\begin{equation}
\mathcal{I}_G=\frac{\hbar}{c}\ell_G. \label{eq:IG}
\end{equation}

Analogous considerations apply to the weak interaction of $\beta$-decay, that was first predicted by Fermi back in 1934. The Regge trajectory for $\beta$-decay is
\begin{equation}
J_F=C\frac{M^2}{M_P^2}=\frac{G_F(Mc/\hbar)^2}{\hbar c}, \label{eq:ReggeF}
\end{equation}
where $M_P$ is the proton mass, $C=10^{-6}$, a numerical constant that must be obtained from experiment, and 
\begin{equation}
G_F=\hbar c\left(\frac{10^{-3}\hbar}{M_Pc}\right)^2, \label{eq:GF}
\end{equation}
is the coupling strength of the four-fermion contact interaction. $G_F$ has the dimensions of energy$\times$volume, and the last factor in \eqref{eq:ReggeF} is the square of the inverse of the Compton wavelength for a particle of mass $M$. It is what is required to make $J_F$ dimensionless. 

The effective range of the weak interaction is
\begin{equation}
\ell_F=\surd\left(\frac{G_F}{\hbar c}\right), \label{eq:ellF}
\end{equation}
which is the inverse of the string length, \eqref{eq:ellS}. Consequently, the mass of the weak interaction is
\begin{equation}
M_F=\hbar\Big/\surd\left(\frac{G_Fc}{\hbar}\right)=\surd\left(\frac{\hbar^3}{G_Fc}\right)=10^3M_P. \label{eq:MF}
\end{equation}
Given \eqref{eq:ellF}, the characteristic angular velocity is found to be
\begin{equation}
\omega_F=\surd\left(\frac{\hbar c^3}{G_F}\right). \label{eq:omegaF}
\end{equation}
If we didn't know \eqref{eq:MF}, we could calculate it from the Regge trajectory \eqref{eq:ReggeF} by introducing the definition of angular momentum, $J_F=M_F\ell^2_F\omega_F/\hbar$, where \eqref{eq:ellF} and \eqref{eq:omegaF} are introduced. We would then find \eqref{eq:MF}, which is not the mass of the $W^{\pm}$ bosons. Why?

The interconversion of neutrons into protons are charge-changing reactions, and as such involve the transfer of charged $W^{\pm}$. These \lq charged current\rq\ reactions are in contrast to electromagnetic scattering where the exchanged particle, the photon, is uncharged. Such scattering is a \lq neutral\rq\ current reaction. But, if the two are to be of the same strength, then $G_F$, which has the dimensions of $e^2\times L^2$, must be of the form
\begin{equation}
G_F=g^2\lambda_W^2,\label{eq:GF-bis}
\end{equation}
where $\lambda_W$ is the Compton wavelength of the $W^{\pm}$ bosons. Setting \eqref{eq:GF} equal to \eqref{eq:GF-bis} we find
\begin{equation}
\frac{g^2}{\hbar c}=\left(10^{-3}\frac{M_W}{M_P}\right)^2,\label{eq:g}
\end{equation}
which characterizes the coupling of the $W^{\pm}$ to the fermions, whose masses are $M_W=80\mbox{GeV}/c^2$. 

Introducing \eqref{eq:GF-bis} into \eqref{eq:ReggeF} gives
\begin{equation}
J_F=\frac{g^2}{\hbar c}\left(\frac{M}{M_W}\right)^2,  \label{eq:ReggeF-bis}
\end{equation}
which is analogous to the pion-nucleon coupling constant, $f/\hbar c=(g^2/\hbar c)(M_{\pi}/M_P)^2$, where $f^2/4\pi$ is the pion-nucleon coupling constant of the strong interaction, having dimensions of the fine structure constant.~\cite{DP}

Introducing \eqref{eq:GF-bis} into the Fermi mass, \eqref{eq:MF},  results in
\begin{equation}
M_F=\surd\left(\frac{\hbar c}{g^2}\right)M_W=10^{3}M_P. \label{eq:MF-bis}
\end{equation}
If we assume, along with Salam and Weinberg, that electromagnetic and weak interactions have the same intrinsic coupling of bosons to leptons, i.e. $g\simeq e$, thereby unifying the \lq electroweak interaction,\rq\ we find the mass of the $W^{\pm}$ bosons as
\begin{equation}
M_W=10^{3}M_P\surd J_{\gamma}=0.085\times 938\times10^3\mbox{MeV}=80\mbox{GeV}.\label{eq:MW}
\end{equation}

The rotational inertia of the weak interaction is
\begin{equation}
\mathcal{I}_F=\frac{\hbar^2}{M_Fc^2}=M_F\lambda_F^2. \label{eq:IF}
\end{equation}
This identifies the mass in \eqref{eq:ReggeF-bis} as \eqref{eq:MF-bis}.

Consequently, the relation $J\sim M^2$ appears to be ubiquitous in particle physics and has nothing to do with the behavior of a flexible string.   The factors are shown in the Table. The only place we expect an \lq intermediate particle\rq\ is in the weak interaction due to the presence of a numerical factor in the last column. The introduction of the fine structure constant, \eqref{eq:g}, in analogy with the exchange of a neutral boson between leptons will then reduce $M_F$ to $\surd J_{\gamma}M_F=M_Z$ since the orders of magnitude of $W^{\pm}$ and $Z^0$ are the same.

\begin{table}
\begin{tabular}{|llllll|} \hline
\textit{name} \hspace{100pt}& $M_X$\hspace{50pt} & $\ell_X$\hspace{50pt} & $\omega_X$\hspace{50pt} & $\mathcal{I}_X$ \hspace{50pt}&\hspace{50pt} $J_X$\\ \hline\\
\textit{string}\hspace{10pt} $X=S$ & $\surd\left(\frac{\varepsilon\hbar}{c^3}\right)$ & $\surd\left(\frac{\hbar c}{\varepsilon}\right)$ & $\surd\left(\frac{\varepsilon c}{\hbar}\right)$ & $\frac{\hbar}{\omega_S}$ &\hspace{50pt} $M^2/M_S^2$\\ \hline\\
\textit{gravity} \hspace{10pt} $X=G$& $\surd\left(\frac{\hbar c}{G}\right)$ & $\surd\left(\frac{G\hbar}{c^3}\right)$ &$\surd\left(\frac{c^5}{G\hbar}\right)$ & $\frac{\hbar}{\omega_G}$ & \hspace{50pt}$M^2/M_G^2$\\ \hline \\
\textit{weak interaction}\hspace{10pt} $X=F$ & $\surd\left(\frac{\hbar^3}{G_Fc}\right)$ & $\surd\left(\frac{G_F}{\hbar c}\right)$ & $\surd\left(\frac{\hbar c^3}{G_F}\right)$ & $\frac{\hbar}{\omega_F}$ &\hspace{50pt} $10^{-6}M^2/M_P^2$\\ \hline
\end{tabular}
\end{table}

\section{Dynamic Phenomena Related to $J\sim M^2$}

The last column of the Table establishes a dynamic equilibrium between the repulsive centrifugal force and attractive forces. The best known is that of gravity, where the relativistic virial is
\begin{equation}
\frac{v^2}{c^2}=\frac{2GM}{c^2\lambda}=\frac{\alpha}{\lambda},\label{eq:gravity}
\end{equation}
with $\lambda=c/\omega$, and, $\alpha$ is the Schwarzschild radius, $2GM/c^2$. Since we are considering the volume per  particle, we can interchange the radius, $r$, and the wavelength, $\lambda$. Then dividing through by $r$ in \eqref{eq:gravity} it says that the centrifugal forces just balance the attractive force by a central mass $M$ due to gravity. If, instead of the mass, the density, $\varrho=M/r^3$, is constant, \eqref{eq:gravity} reduces to
\begin{equation}
\omega_{\rm{rot}}=v/\lambda=1/\surd(G\varrho). \label{eq:gravity-bis}
\end{equation}
This is the maximum frequency that a star can rotate without its matter being torn off its surface by the centrifugal force.~\cite{Sexl}

According to the last column for the weak interaction in the Table, the mass of the intermediate particle would be given by
\begin{equation}
M_W=10^3M_P\frac{v}{c}. \label{eq:weak}
\end{equation}
Comparing \eqref{eq:MW} with \eqref{eq:weak} we find
\begin{equation}
\frac{v^2}{c^2}=J_\gamma\ll1, \label{eq:weak-bis}
\end{equation}
and which indicates that the electroweak interaction is a non-relativistic effect. Although $\beta$-decay can be relativistic, \eqref{eq:weak-bis} says that the relation $J\sim M^2$ holds only in the non-relativistic regime. It is well-known that Regge poles characterize \lq soft-scattering\rq\ processes where the created particles carry away small momenta in the direction normal to the beam. Since Regge theory has little to do with \lq hard-scattering\rq\ processes, we would not expect such trajectories exist for such processes. However, \eqref{eq:MF} implies another mass, which is $11.7$ times that of $M_W$. For such a particle we would, again, expect $J\sim M^2$ to hold.

Finally, for the \lq string,\rq\ we have 
\begin{equation}
\frac{v^2}{c^2}=\frac{M}{M_S}.\label{eq:string}
\end{equation}
If the string has constant density, $\varrho=M/\lambda$, \eqref{eq:string} becomes
\begin{equation}
\frac{v}{c}=\surd\left(\frac{\varrho c^2}{\varepsilon}\right)=\frac{r}{r_0}. \label{eq:string-bis}
\end{equation}
In general, at intermediate points along the string, $\varepsilon>\varrho c^2$, and only at the ends will there be an equality. In fact, \eqref{eq:omegaS-bis} would predict superluminal velocities independent of where one is on the string. The expression for the mass of a string, \eqref{eq:MS}, shows that the higher the energy density, the larger the mass. The Regge trajectory, \eqref{eq:Regge}, shows that the higher the energy density, the smaller will be its angular momentum. This is in glaring contrast to the transverse vibrations of a string where the higher the tension, or the lighter the string, the larger will be the velocity of propagation.~\cite[p. 75]{BZ}

\section{Cosmic Strings are Out of this World}

Because a string has mass, one would be inclined to believe that there will be gravitational attraction. This is what Newtonian theory tells us, but not Einstein's relativistic theory  for the tension of a string would exert a \lq negative\rq\ gravitational attraction that would cancel precisely the positive gravitational attraction of its mass. Yet, while strings exert no gravitational attraction, they will affect the geometry of planes perpendicular to the string.~\cite{BZ} 

Suppose we revolve around a string at a fixed distance $r$. Surprisingly, we would find that the circumference of our circle is not $2\pi r$, but~\cite{BZ}
\begin{equation}
C=\left(2\pi-\Delta\right)r, \label{eq:circ}
\end{equation}
where $\Delta$ is the \lq deficit angle.\rq\ According to general relativity it is
\begin{equation}
\Delta=\frac{8\pi G\varepsilon}{c^4}=\frac{8\pi G\varrho}{c^2}. \label{eq:Delta}
\end{equation}
Expression \eqref{eq:Delta} looks odd because instead of having one parameter measuring the strength of the coupling we now have two. 

If the cosmic string is placed between a source of light and an observer, it can produce gravitational lensing, where light avails itself of more than one geodesic to reach the observer. Such compact objects can thus produce more than two images, and since the geometry is curved, the images may appear distorted, and could be quite different. This, according to string theorists, would provide firm evidence for the existence of cosmic strings.

Apart, from our earlier objection to \eqref{eq:Delta}, \eqref{eq:circ} can't be right. According to Fermat's principle of least time, the optical path, 
\begin{equation}
I=c\tau=\int\eta\surd(2T)\;dt, \label{eq:Fermat}
\end{equation}
connecting any two arbitrary points, should be stationary. $T$ is the kinetic energy per unit mass, and it is related to the Euclidean metric according to
\begin{equation}
\surd(2T)\;dt=ds=\surd\left(dr^2+r^2d\varphi^2\right), \label{eq:KE}
\end{equation}
in polar coordinates. 

Rather, if the medium is inhomogeneous, the index of refraction, $\eta$, will vary with $r$, but not with $\varphi$. If light is propagating in a gravitational field, the optical path length is~\cite{BHL}
\begin{equation}
I=\int\eta ds=\int ds^{\prime}=\int\surd\left(1+\frac{2\alpha}{r}\right)\;ds.\label{eq:Fermat-bis}
\end{equation}
Moreover, if we consider $r=\mbox{const.}$, then \eqref{eq:Fermat-bis} will reduce to
\begin{equation}
C=\int_0^{2\pi}ds^{\prime}=r\int_0^{2\pi}\surd\left(1+\frac{2\alpha}{r}\right)\;d\varphi=\surd\left(1+\frac{2\alpha}{r}\right)2\pi r. \label{eq:circ-bis}
\end{equation}
For weak fields, \eqref{eq:circ-bis} would be approximated by
\begin{equation}
C=\left(1+\frac{2GM}{r}\right)2\pi r. \label{eq:circ-tris}
\end{equation}

So, instead of having an angle \lq deficit\rq, we have an angle \lq excess.\rq\ This is none other than Einstein's old result of the circumference of a uniformly rotating disc~\cite{BHL}. Einstein reasoned, incorrectly, that the rulers lined up along the circumference of the disc would undergo a FitzGerald-Lorentz contraction, so more would be needed when the disc is in motion than when it was at rest. This is fallacious since the circumference too would likewise undergo the same contraction. Expression \eqref{eq:circ-tris} would also allow Einstein to conclude that a gravitational field mimics uniform acceleration so that all forms of acceleration are equivalent. This he called his \lq equivalence principle.\rq\ This also is inaccurate since the gravitational field appears as a cause for inhomogeneity in the expression for the index of refraction, \eqref{eq:Fermat-bis}, and, in its absence, the disc would still rotate with unform acceleration.

If the latter is true then we would expect that the circumference is still $2\pi r$. That would be if we are in Euclidean space. Rather, in hyperbolic space the metric is not given by \eqref{eq:KE}, but, rather by~\cite[p. 487]{BHL}
\begin{equation}
d\bar{s}^2=d\bar{r}^2+R^2\sinh^2(\bar{r}/R)d\varphi^2, \label{eq:Beltrami}
\end{equation}
where $\bar{r}=R\tanh^{-1}{r/R}$ is the straight-line segment in hyperbolic space, and $R$ is the radius of curvature. A common, and almost universal, error is to set $r=R\sinh\bar{r}/R$.~\cite{LL} While it is true that the surface area is $4\pi R^2\sinh^2\bar{r}/R$, it is so for a different reason, i.e., the effect of distortion has not been taken into account.~\cite{HB} This is related to the fact that the sum of the angles of a triangle is inferior to $\pi$.

At constant $r$, the Beltrami metric, \eqref{eq:Beltrami} gives a circumference
\begin{equation}
C=\int_0^{2\pi}d\bar{s}=\frac{r\;d\varphi}{\surd\left(1-r^2\omega^2/c^2\right)}=\frac{2\pi r}{\surd\left(1-r^2\omega^2/c^2\right)}, \label{eq:Einstein}
\end{equation}
which is Einstein's old result~\cite{CM}, where we have set the radius of curvature, $R=c/\omega$. As the derivation shows, it has been obtained from the wrong reasons.~\cite{BHL} Thus, all forms of acceleration are not equivalent. The centrifugal potential is already included in the Euclidean metric, $\dot{r}^2+L^2/r^2$, where
$L=r^2\dot{\varphi}$ is the angular momentum per unit mass. Hyperbolic space modifies the result   $L=\mbox{const.}$ to read
\begin{equation}
\frac{L^2}{r^2\left(1-r^2\omega^2/R^2\right)}=\mbox{const.} \label{eq:L}
\end{equation}
The same result is obtained in the general theory of relativity~\cite[p. 349 formula (18)]{CM}, provoking the comment that \eqref{eq:L} \lq\lq cannot in general be interpreted as angular momentum, since the notion of a \lq radial vector\rq\ occurring in the definition of the angular momentum has an unambiguous meaning in a Euclidean space.\rq\rq We  strongly disagree with M\o ller. The effect here is the destruction of the uniformity of space by a static gravitational field, and has the same effect  in varying the index of refraction in air which decreases with altitude.

Cosmic strings are supposedly left-overs from the early universe, and can stretch across the observable universe~\cite{BZ}. If they are to be detected by gravitational lensing then string theorists should be looking for angular excess instead of deficit.

\section{Black Holes in Anti-de Sitter Universes of $5$ Dimensions}

String theory needs \lq supersymmetry\rq\ to introduce fermions which also lowers the dimensions from 26 to 10. Then starting with $10$ dimensional super string theory and compactifying six of those dimensions will result in a $4$ dimensional super symmetric universe. However, this will not cover the case of black holes which supposedly require $5$ dimensions in an anti-de Sitter (AdS) universe. This is indeed intriguing for it not only reproduces results which are blatantly incorrect, it needs $5$ dimensions to do so! Black holes supposedly radiate thermal energy with a black body spectrum. String theorists claim that $5$ dimensions are required, but consider black body radiation in only $3$ dimensions.

The good news, string theorists tell us, is that AdS supports black holes in $5$ dimensions. The Schwarzschild metric~\cite[p. 325 formula (79)]{CM}
\begin{equation}
ds^2=-\left(1-\frac{\alpha}{r}-\lambda r^2/3\right)dt^2+\left(1-\frac{\alpha}{r}-\lambda r^2/3\right)^{-1}dr^2+r^2\left(\sin^2\vartheta d\varphi+d\vartheta^2\right), \label{eq:Schwarz}
\end{equation}
where $\lambda$ is a small, positive, constant, gets mutilated into~\cite[p. 554 formula (23.104)]{BZ}
\begin{equation}
ds^2=-\left(1+\frac{r^2}{R^2}- \frac{r^2_0}{r^2}\right)dt^2+\left(1+\frac{r^2}{R^2}- \frac{r^2_0}{r^2}\right)^{-1}dr^2+r^2d\Omega_3^2, \label{eq:Schwarz-bis}
\end{equation}
by the string theorists. \lq\lq The parameter $r_0$ tells us that we have a black hole\rq\rq\ in an AdS of $5$ dimensions, where $d\Omega_3^2$ \lq\lq denotes the metric of a $3$-sphere $\mathbb{S}^3$ of unit radius.\rq\rq\ Moreover, \lq\lq one can write $r_0$ in terms of the mass $M$ of the hole and the $5$ dimensional Newton's constant: $r_0^2\sim G^{(5)}M$.\rq\rq~\cite[p. 555]{BZ}

String theorists also tell us that the Schwarzschild radius, $r_{+}$ is that value of $r$ that nullifies the coefficient of $dt^2$ in the metric \eqref{eq:Schwarz-bis}. It will only be proportional to $r_0$ in the limit $r\ll R$. In general,
\[\frac{r_{+}}{r_0}=\frac{1}{1+r_{+}^2/R^2}<1,\]
which makes the dubious claim that the Schwarzschild radius is always less than $r_0$, which is proportional to the mass $M$ of the black hole. Moreover, $R$, the radius of curvature, is supposedly not a limit to the size of a black hole: Small black holes are those for which $r_{+}\ll R$, while large black holes have $r_{+}\gg R$.

Now enters thermodynamics: For any metric of the form
\[ds^2=-f(r)dt^2+f(r)^{-1}dr^2+\cdots,\]
the Hawking temperature is defined by, $T=f^{\prime}(r_{+})/4\pi$~\cite[p. 556]{BZ}, where the prime stands for the differentiation with respect to $r$,  
\begin{equation}
T_H=\frac{R^2+2r_{+}^2}{2\pi r_{+}R^2}, \label{eq:Hawking}
\end{equation}
which has been evaluated at $r=r{+}$.
Consequently, small black holes will have a Hawking temperature of
\begin{equation}
T=\frac{1}{2\pi r_{+}}, \label{eq:piccolo}
\end{equation}
while for large black holes,
\begin{equation}
T=\frac{r_{+}}{\pi R^2}. \label{eq:grande}
\end{equation}
String theorists even readily admit that \lq\lq This is unusual: once a black hole is large enough its temperature grows with its size.\rq\rq~\cite[p. 556]{BZ} Nonetheless, \lq\lq [t]his is an important qualitative feature of black holes.\rq\rq\ 

\begin{figure}
	\centering
		\includegraphics[width=0.5\textwidth]{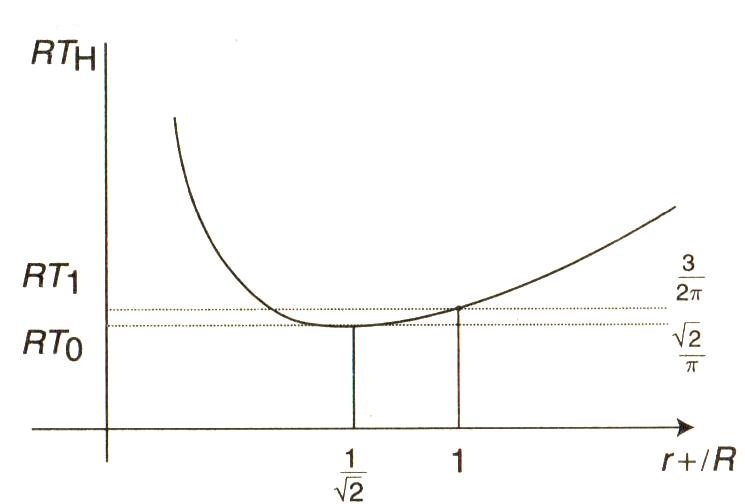}
	\caption{This figure is taken from~\cite{BZ} which plots the product of the Hawking temperature and the AdS radius versus the Schwarzschild radius $r_{+}$ scaled by the same radius. For each value of the Hawking temperature above $T_0$ there exist two black holes.  For temperatures less than $T_0$ there cannot be \lq\lq a black hole dual since there is no black holq\rq\rq}
	\label{fig:BZ}
\end{figure}

For a fixed radius of curvature, $R$, we are told that the Hawking temperature has a lower bound given by
\[T_H\ge T_0=\frac{\surd 2}{\pi R},\]
where the lower bound, $T_0$, is attained at $r_{+}/R=1/\surd 2$. This is depicted in Fig.~\ref{fig:BZ}. At temperature $T_1$ there occurs a \lq\lq Hawking-Page transition.\rq\rq\ 

Temperature may mean different things to different people. But, all would agree that temperature is a \emph{monotonic\/} function of whatever it is a function of. If the universe is expanding into a void, then the adiabatic expansion should show a continual decrease in temperature as the radius of the sphere increases, $TV^{1/3}=\mbox{const}$. We certainly should not expect something like \eqref{eq:piccolo} and \eqref{eq:grande} occurring in the same system under the same conditions. 

Nonetheless, it is absolutely necessary in order to show that black holes radiate with a black body spectrum.  The last term in the mutilated metric, \eqref{eq:Schwarz-bis} shows that the horizon is a three-sphere, $\mathbb{S}^3$, of radius $r_{+}$, although previously we were told that it was for a unit radius.  The metric, \eqref{eq:Schwarz-bis}, for fixed $r$ becomes
\begin{equation}
ds^2\simeq\frac{r^2}{R^2}\left[-dt^2+R^2d\Omega_3^2\right].\label{eq:BZ}
\end{equation}
The following conclusions are \emph{non-seguiturs\/}~\cite[pp. 555-556]{BZ}:
\begin{itemize}
\item \lq\lq While the radius, $R$, is immaterial in the zero-temperature case, it is not immaterial now.\rq\rq\
\item \lq\lq The radius $R$ is the radius of the sphere where the field theory at temperature $T_H$ lives.\rq\rq\
\item \lq\lq One cannot rescale with impunity the time coordinate $t$ in \eqref{eq:BZ} because at non-zero temperature the time coordinate carries information about temperature.\rq\rq\
\item \lq\lq The product $RT_H$ of the radius and the temperature is the \emph{only\/} invariant information.\rq\rq\
\end{itemize}

A generalization of the Bekenstein expression for the black hole entropy says that the black hole entropy is the ratio of the \lq area\rq\ of the event horizon divided by the $5$-dimensional Newtonian constant, $G^{(5)}$:
\begin{equation}
S_{BH}=\frac{A_{\rm{hor}}}{4G^{(5)}}. \label{eq:S}
\end{equation}
Now comes the punchline: The last term in the metric \eqref{eq:Schwarz-bis} shows that the horizon is a $3$-sphere of radius $r_{+}$ so the area is $A_{\rm{hor}}=2\pi^2r_{+}^3$, and not proportional to the square of the radius as it would be had the field theory lived in real $3$-dimensional space. Then from \eqref{eq:grande} we see that $r_{+}$ is proportional to the temperature $T_H$ so that the cube of one will give the cube of the other. And since we \emph{know\/} that the entropy of black body radiation is proportional to the cube of the temperature, it would seemingly make compatible the entropy of a black hole with the entropy of black radiation. 

The whole thing is a \emph{mise en sc\`ene\/}, and even if we didn't know what the gravitational constant looks like in $5$-dimensions we could find it by comparing \eqref{eq:S} with the entropy of black body radiation. This would give $G^{(5)}=2\pi R^3/N^2$, and the resulting entropy would be
\begin{equation}
S_{BH}=\frac{2\pi^2r_{+}^3}{2\pi R^3/N^2}=\frac{N^22\pi^2R^3r_{+}^3}{2\pi R^6}=\half\pi^2N^2T_H^3\cdot(2\pi R^3).\label{eq:S-bis}
\end{equation}
{\emph{The radiation formula \eqref{eq:S-bis} applies only to black holes on the ascending branch in Fig.~\ref{fig:BZ}, so that small black holes on the descending branch of the curve would either not radiate, or radiate with some other spectrum than a black body\/}. But Hawking~\cite{SH} found precisely that small black holes with temperature given by \eqref{eq:piccolo} do radiate with a thermal spectrum.

However, it was previously asked how the horizon size $R$ is determined by the number of \lq branes,\rq\ so the two appear to be related.~\cite[p. 542]{BZ} It is also ironical to see that while string theorists hold to black body radiations in 3 dimensions, they claim that 5 dimensions are needed. Long ago, Lord Rayleigh~\cite{LR} gave a generalization to Stefan's law in the case the spatial dimension would be different than 3. By generalizing the equation of state between the pressure $p$ and internal energy density $\varepsilon$, $p=\third\varepsilon$ to $p=\varepsilon/\eta$, where $\eta>3$ for dimensions higher than $3$, Rayleigh generalized Stefan's law to $\varepsilon\sim T^{1+\eta}$, and, consequently, the entropy density varies as $s\sim T^{\eta}$. So one cannot retain black body radiation in $3$ dimensions, \eqref{eq:S-bis}, while, at the same time, requiring higher spatial dimensions.

\section{Regge Scattering Amplitudes and Automorphic Functions}
An automorphic function is an extension of the concept of a periodic function to the more general case of discontinuous groups. That is, circular functions are automorphic with respect to the group $\{2\pi n|n\in\mathbb{Z}\}$, and the elliptic functions are automorphic with respect to the group $\{m\omega_1+n\omega_2|m,n\in\mathbb{Z}\}$, with $\omega_1/\omega_2\not\in\mathbb{R}\}$. Circular functions are associated with a sphere, a solid body of genus 0. It has zero moduli and consists in transforming a sphere in an infinite number of ways. Elliptic functions are of genus $g=1$ with one modulus, the ratio $\omega_1/\omega_2$. For $g>1$ there will be $3g-3$ moduli. Moduli are those constants which play the role of invariants in a uniform transformation from the $z$- to the $\omega$- complex planes under the mapping $f(\omega,z)=0$. For elliptic functions, the $\omega$-plane is tessellated by parallelograms.

All this was known before 1880. The modular function, $f$, defined on the half-plane $\mbox{Im}(z)>0$ was known to be automorphic with respect to the modular group of transformations,
\[z\longmapsto\frac{az+b}{cz+d},\]
where $a, b, c, d\in\mathbb{Z}$ and $ad-bc=1$. If the coefficients satisfy these conditions then
\[f\left(\frac{az+b}{cz+d}\right)=f(z).\]
The modular group arose from considerations about the equivalence of birational quadratic forms, and was known already to Gauss in this context. The generators of the modular group are the translations and inversions,
\begin{gather*}
T_t=\begin{pmatrix} 1 & 1 \\ 0 & 1\end{pmatrix},\qquad\qquad\mbox{and}\qquad\qquad T_i= \begin{pmatrix} 0 & 1\\ -1 & 0\end{pmatrix},\end{gather*}
respectively.

 The Riemann space is one of triangular tessellations where the angles of the triangle are related to the exponents of the indicial equations about the singular points. Poincar\'e discovered that the automorphic functions are periodic functions with respect to the class of all fractional linear transformations.

Why string theory got started at all was Veneziano's~\cite{GV} Euler beta integral for the expression of the relativistic four point scattering amplitude according to Regge theory. The Regge trajectories determine  the exponents in the beta integral, and the mapping was from the half-plane of poles located on the real axis of the $z$-plane to the Riemann $\omega$-plane comprised of non-overlapping triangles that tessellate the disc completely. The poles in the $z$-plane determine the angles of the triangles in the $\omega$-plane.

As an illustration, we will derive the leading term in the scattering amplitude of four particles $a+b\longrightarrow c+d$, known as the Regge limit. Following convention, we introduce the Mandelstam variables
\begin{gather*}
s=-(p_a+p_b)^2=-(p_c+p_d)^2,\\
t=-(p_a-p_c)^2=-(p_b-p_d)^2,
\end{gather*}
which apart from sign, are the total energy and momentum transfer, respectively. By defining the crossed momentum transfer, $u=-(p_a-p_d)^2$, it is easy to see that
the sum of the Mandelstam variables, 
\[s+t+u=4M^2,\]
gives four times the square of the mass of each particle if the particles have the same mass. We will consider the momentum transfer, $t$, to be fixed, and let the total energy, $s$ to be variable. It will have branch points at $s=s_0$ and $s=\infty$. This corresponds to a single particle threshold. Mapping the $s$-plane onto the $z$-plane by $z=s-s_0$, at $s=s_0$, $z=0$, and at $s=\infty$, $z=\infty$. In the $\omega$-plane this will map to two circular arcs which cut each other at the same angle $\lambda\pi$ to form a crescent as shown in Fig.~\ref{fig:banana}.
 
\begin{figure}
	\centering
		\includegraphics[width=0.3\textwidth]{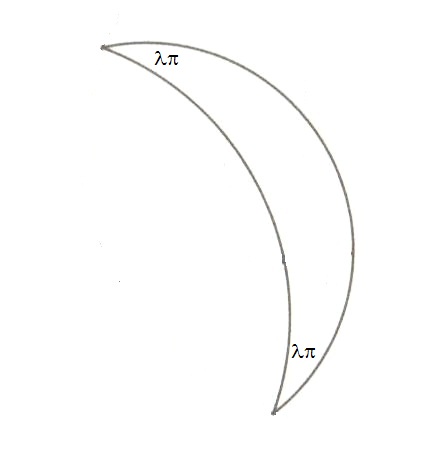}
	\caption{Two circular arcs cut each other at an angle $\lambda\pi$ to form a crescent.}
	\label{fig:banana}
\end{figure}
 Fuch's great discovery was that the hypergeometric equation is the only differential equation, of order greater than one, for which the exponents of the two solutions in the immediate vicinity of the poles uniquely determine the angles of the triangle. In the present case it is given by
\begin{equation}
\frac{d^2w}{dz^2}-\frac{1-\lambda^2}{z^2}w=0, \label{eq:hyper}
\end{equation}
since the angle at infinity never explicitly appears.

To motivate the physical interpretation of \eqref{eq:hyper}, consider the Schr\"odinger equation at small radial values, $z$, for which only the centrifugal barrier dominates,
\begin{equation}
\frac{d^2\psi}{dz^2}-\frac{\ell(\ell+1)}{z^2}\psi=0, \label{eq:Schrodinger}
\end{equation}
where $\ell$ is the angular momentum. Now, \eqref{eq:Schrodinger} can be cast in the form \eqref{eq:hyper} by writing
\begin{equation}
\lambda=2\ell+1. \label{eq:lambda}
\end{equation}
But, the condition that the angle, $\lambda\pi$ be less than $\pi$ implies
\begin{equation}
0\ge\ell\ge-\frac12. \label{eq:ell}
\end{equation}
In this case,  $P_{\ell}(\cos\vartheta)$ will no longer be a Legendre polynomial, but, rather, a hypergeometric function with a branch cut going from $-1$ to $-\infty$.

The two independent solutions to the equation \eqref{eq:hyper} are
\begin{equation}
w_1=z^{1+\ell} \hspace{30pt}\mbox{and}\hspace{30pt} w_2=z^{-\ell}. \label{eq:soln}
\end{equation}
Consider now the quotient of the solutions
\begin{equation}
\omega=\frac{w_1}{w_2}=\lambda\int_0^{z}z^{\lambda-1}dz. \label{eq:quotient}
\end{equation}
The second equality shows that $\omega$ is the Schwarz-Christoffel mapping function of part of the Veneziano amplitude for the four point elastic scattering amplitude, where $\lambda-1$ would be an exterior angle of one of the vertices.\footnote{This has been done in~\cite{AC}. In order to do so it was necessary to identify the angles with $1-\mbox{Re}\;\alpha$, which does not cover the zones of bound states and resonances [cf. Fig.~\ref{fig:trajectory}.} If there was an additional fixed point at $z=1$, \eqref{eq:quotient} would have been an incomplete $\beta$-function, known as a Schwarz triangle function in the $\omega$-plane.~\cite{LA}

The linear fractional transformation,
\begin{equation}
\omega^{\prime}=\frac{a\omega+b}{c\omega+d}, \label{eq:Mob}
\end{equation}
where the coefficients, $a,b,c,\in\mathbb{Z}$, $ad-bc=1$, and $\omega^{\prime}=\omega_1^{\prime}/\omega_2^{\prime}$, will transform two circles that cut one another into two other circles that intersect at the same angles. In other words, the M\"obius transform \eqref{eq:Mob} is conformal. 

 If the $\omega$-circles intersect in $-i$ and $+i$ corresponding to the origin  and the point at infinity in the $z$-plane, then the elliptic substitution,
\begin{equation}
\omega^{\prime}=\frac{\cos\varphi z+\sin\varphi}{-\sin\varphi z+\cos\varphi}, \label{eq:rot}
\end{equation}
where $\varphi=\lambda\pi/2$ is a periodic substitution of period $2/\lambda$ which, by repeated application to the area of the crescent, will divide the plane into $2/\lambda$ regions, that, with the exception of two, have the same crescent form~\cite[p. 630]{ARF}. 

On the other hand, if the vertices of the cresent are at $-1$ and $+1$ in the $\omega$-plane, corresponding to the fixed points $0$ and $\infty$, respectively, the hyperbolic substitution,~\footnote{It is a great pity that Poincar\'e, who developed the theory of automorphic functions,~\cite{HP} did not appreciate that the same type of transforms, later to become known as Lorentz transforms, did not apply to relativity.}
\begin{equation}
\omega^{\prime}=\frac{\cosh\Phi z+\sinh\Phi}{\sinh\Phi z+\cosh\Phi}, \label{eq:stretch}
\end{equation}
where $\Phi=\lambda\pi/2$, will stretch the symmetrically the unit disc $\mathbb{D}$ in the $\omega$-plane away from the source at $-1$ towards the sink at $+1$. Hence, the transform from an elliptic to hyperbolic substitution is tantamount to making the angle imaginary. The trajectories appear as lines of force on the left-hand side of Fig.~\ref{fig:string}, where the lines  start from the positive charge and end on the negative charge. The fundamental region is the crescent with equal angles, $\lambda\pi$.  The lines of electrical force are, therefore, actually tessellations of the hyperbolic plane!

To determine the elastic scattering amplitude, we observe that when $t$ is small and $s\rightarrow\infty$, $\cos\vartheta\propto-s$, and it can be shown that~\cite{PC}
\begin{equation}
P_{\ell}(-\cos\vartheta)\propto s^{\ell}=e^{\ell(\ln s)}\hspace{30pt}\mbox{for}\hspace{30pt} \mbox{Re}\;\ell>-\half.\label{eq:Legendre}
\end{equation}
Since \eqref{eq:Legendre} is proportional to the scattering amplitude, In order for there to be an exponential falloff in the elastic scattering amplitude with increasing momentum transfer,  $\ell=\alpha(t)<0$.   This together with the condition in \eqref{eq:Legendre} is precisely the interval \eqref{eq:ell}.

A Regge trajectory takes the linear form
\begin{equation}
\ell=\alpha(t)=\alpha^{\prime}t+\alpha_0, \label{eq:trajectory}
\end{equation}
as shown in Fig.~\ref{fig:trajectory}
\begin{figure}
	\centering
		\includegraphics[width=0.5\textwidth]{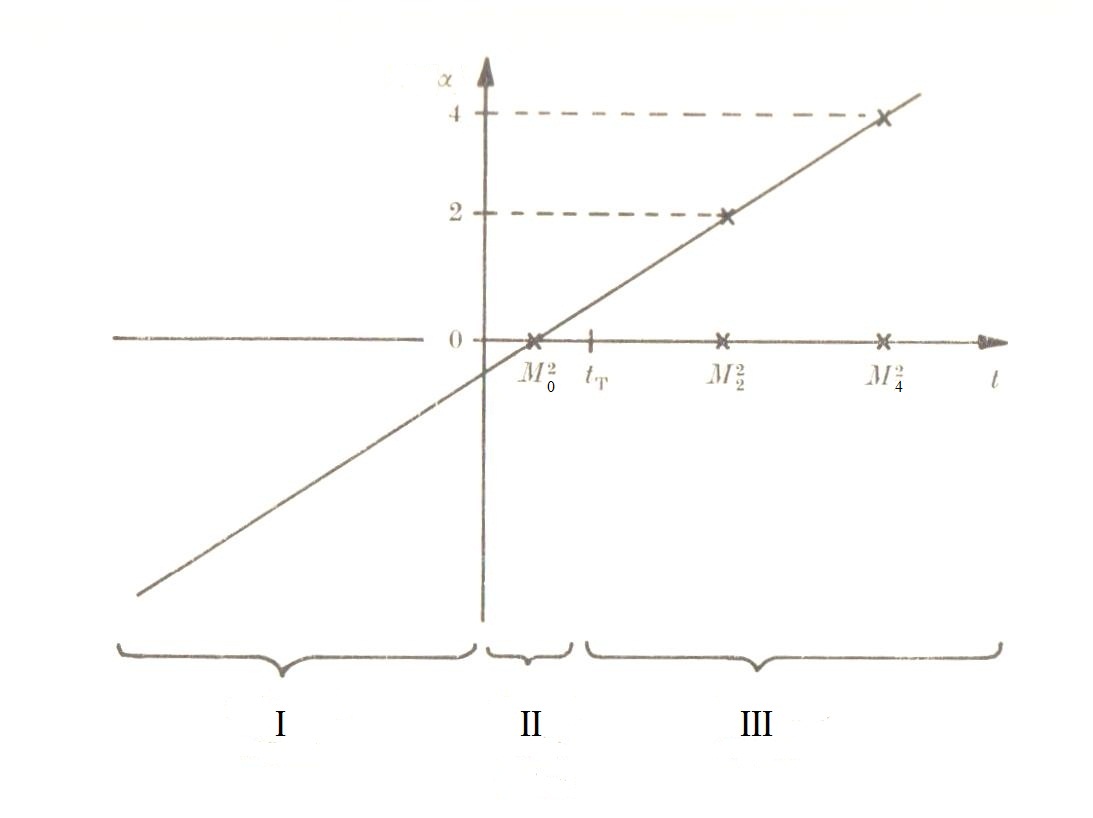}
	\caption{A Regge trajectory with even signature spanning three regions: I s-channel where s has the meaning of total energy, II the bound state region, and III the resonance region.}
	\label{fig:trajectory}
\end{figure}
For negative values of \eqref{eq:trajectory}, $t$ has the significance of momentum transfer and $s$ of total energy. Thus, \eqref{eq:trajectory} governs high energy behavior in the $s$-channel. The trajectory contributes to the power behavior, \eqref{eq:Legendre}, of the $s$-channel amplitude. In this region $\alpha(t)$ can be detected for all values of $t<0$.

As $t$ crosses into the positive region, it takes on the meaning of total energy, and $\alpha(t)$ becomes observable only at integers differing by values of $2$ (even signature), as shown in Fig.~\ref{fig:trajectory}. The values of $t$ that make $\alpha(t)$ pass through non-negative integers $\ell$ correspond to bound states (II), or resonances (III), of angular momentum $\ell$,
\begin{equation}
t=M_{\ell}^2=\frac{\ell-\alpha_0}{\alpha^{\prime}}, \hspace{30pt} \ell=0,2,4,\ldots \hspace{30pt} (\alpha_0<0). \label{eq:M}
\end{equation}
For $t<t_T$ (the threshold), $\ell=0$ corresponds to a bound state of imaginary mass $M_0$ having zero spin, i.e. a tachyon. Moreover, there are resonances of masses $M_2$ and $M_4$ having spins $2$ and $4$, respectively.

\emph{It is therefore apparent that automorphic functions are related to continuous, negative, values of $\ell$, while discrete, positive, values of $\ell$ violate the condition that the angle \eqref{eq:lambda} be less than $1$\/.} That is, there is a discrete fundamental region for a continuous range of values of $\ell$ given by \eqref{eq:ell}.

If the total cross-section is to be nearly constant at high $s$, it needs $\alpha_0=1$, which is the limit for the strong interaction under crossing~\cite[p. 183]{PC}. This trajectory is called a Pomeron, and was used to account for the asymptotic behavior of the total cross-section. This is precisely the trajectory the string theorists use, and it would make the zero angular momentum state a  tachyon according to \eqref{eq:M}. Since they do not impose changes in angular momentum in intervals of $2$, the $\ell=1$ would be massless~\cite[p. 626]{BZ}. Particles of new masses, $M_{\ell}$, would appear for ever increasing non-negative integer values of the angular momentum. However, all this has nothing whatsoever to do with a vibrating flexible string.

\end{document}